# Impact of COVID-19 on Forecasting Stock Prices: An Integration of Stationary Wavelet Transform and Bidirectional Long Short-Term Memory

Daniel Štifanić[1], Jelena Musulin[2], Adrijana Miočević[3], Sandi Baressi Šegota[4], Roman Šubić[5], Zlatan Car[6]

[1]dstifanic@riteh.hr

[2]jmusulin@riteh.hr

[3]adrijana.miocevic@uniri.hr

[4]sbaressisegota@riteh.hr

[5]roman.subic@hnb.hr

[6]car@riteh.hr

[1,2,4,6]Faculty of Engineering Rijeka, University of Rijeka; Vukovarska 58, 51000 Rijeka, Croatia

[3]University of Rijeka; Trg braće Mažuranića 10, 51000 Rijeka, Croatia

[5]Croatian National Bank; Trg hrvatskih velikana 3, 10000 Zagreb, Croatia

**Abstract**

COVID-19 is an infectious disease that mostly affects the respiratory system. At the time of this research being performed, there were more than 1.4 million cases of COVID-19, and one of the biggest anxieties is not just our health, but our livelihoods, too. In this research, authors investigate the impact of COVID-19 on the global economy, more specifically, the impact of COVID-19 on financial movement of Crude Oil price and three U.S. stock indexes: DJI, S&P 500 and NASDAQ Composite. The proposed system for predicting commodity and stock prices integrates the Stationary Wavelet Transform (SWT) and Bidirectional Long Short-Term Memory (BDLSTM) networks. Firstly, SWT is used to decompose the data into approximation and detail coefficients. After decomposition, data of Crude Oil price and stock market indexes along with COVID-19 confirmed cases were used as input variables for future price movement forecasting. As a result, the proposed system BDLSTM+WT-ADA achieved satisfactory results in terms of five-day Crude Oil price forecast.

# 1. Introduction

Infectious diseases have always been a threat to humanity, especially those about which little or nothing is known. World Health Organization (WHO) describes pandemic as "the worldwide spread of a new disease" and although in such times the greatest concern is how to save human lives, the first following objective is how to save the economy and preserve the well-being [1]. In recent history, it is possible to observe the impact of Spanish flu (1918-1919) on the economy. According to Centers for Disease Control and Prevention (CDC) estimates, roughly 500 million people were taken ill with the disease, which ultimately took the lives of about 50 million worldwide [2]. Even though the economic data from the early 20th century is rare, it has been noted that the impact of business closures has led to unemployment, and businesses that have survived have suffered huge losses. The comparison can be drawn with the pandemic from the recent past, too. During the 2003 SARS (Severe Acute Respiratory Syndrome), which lasted less than a year, business saw enormous revenue plunge. Similar scenario happened in 2009 when expansion the H1N1 flu triggered numerous consequences [3][4]. Pandemic like COVID-19 will surely have a significant influence on the global economy, as well as impact on the financial markets. From 24 to 28 February 2020, stock markets worldwide reported their largest one-week declines since the 2008 financial crisis. Traders began to sell shares out of fear, and as a result, a market-wide circuit breaker was triggered four times in March [5][6]. The breaks were made for 15 minutes each in the hope that the situation would calm down. Every pandemic is unique and it is unlikely to expect the same results, but direction and movement can be predicted which is important for a timely response. A recent occurrence of pandemic has created a supply and a demand shock which is significantly different in comparison with other crises. Starting with the supply-side reductions due to the astonishing closures of factories and labor shortages, the global economy was simultaneously affected by the demand-side shock with immediate reduction in consumer spending. These shocks have ultimately resulted in shifting aggregate supply and aggregate demand downward and, consequently, in reducing national and global gross domestic products.

Forecasting stock prices has always been considered a challenging task due to the fact that stock market tends to be non-stationary, non-linear and highly noisy [7]. Artificial Intelligence (AI) algorithms have been proven successful in solving problems such as predicting stock prices [8] as well as other various fields of science, technology and medicine [9-11]. Numerous factors influence financial market performance, and even financial experts find it complicated to make accurate predictions. The algorithm that may be efficient in commodity and stock market forecasting is a Bidirectional Long Short-Term Memory (BDLSTM) [12]. This algorithm is a combination of Bidirectional Recurrent Network (BDRNN) and Long Short-Term Memory (LSTM) cells. Such combination causes the BDLSTM to have the advantage of LSTM with feedback for the next layer [13].

Althelaya et al. (2018) demonstrate the use of BDLSTM for the most challenging real-world application for time-series prediction [14]. Jia et al. (2019) show the use of Bidirectional LSTM to predict the accuracy of GREE stock price, and achieve good results [15]. Eapen et al. (2019) offer a view into a combination of multiple pipelines of convolutional neural network and bidirectional long short term memory units and its use for stock market index prediction [16].

In order to decompose high complexity data of commodity and stock market indexes and in the same time retain translation invariance, Stationary Wavelet Transform (SWT) was utilized. Since SWT is shift-invariant and non-decimated, it can be used for feature extraction, change detection and pattern recognition. SWT can be described as follows: at each level, after the signal is convolved with high-pass and low-pass filters, resulting sequences have the same number of samples as the original signal [17].

Bai et al. (2016) demonstrate the successful use of SWT and backpropagation neural network (BPNN) to forecast daily air pollutants concentrations and the results show that the SWT-BPNN model has better forecasting performance for the three air pollutants than BPNN model without SWT [18]. Supratid et al. (2017) show development of a reservoir inflow integrated forecasting model, relying on SWT and nonlinear autoregressive neural network with exogenous input (NARX), and achieve good results with relatively accurate predictions [19].

Three major U.S. indexes: Dow Jones Industrial Average, S&P 500 and NASDAQ Composite along with Crude Oil price, are chosen as the research objects. Sample data is selected from March 22, 2000 to April 7, 2020. Wang et al. (2012) show the cross-correlations between Crude Oil market and Dow Jones Industrial Average, S&P 500 and NASDAQ Composite stock market from the perspective of econophysics, and they found that cross-correlated behavior between Crude Oil market and other three U.S. stock market is nonlinear and multifractal [20].

Datasets for each stock market index (Dow Jones Industrial Average, S&P 500 and NASDAQ Composite) along with Crude Oil price were obtained from Yahoo finance website [21] while the data of COVID-19 confirmed cases was obtained from the Johns Hopkins University Center for Systems Science and Engineering (JHU CSSE) [22]. At the time, when this research was performed, data of commodity and each stock market index consisted of 4992 data-points, which were split into the training and testing sets.

The aim of this research is to integrate SWT with BDLSTM in order to predict the movement of aforementioned commodity and stock market indexes during the COVID-19 outbreak.

COVID-19 caused huge shock to the global economy including commodity prices as well as stock market [23]. With implementation and forecasting price movement, it is expected to make a prompt and significant contribution in terms of understanding and responding to impact of COVID-19 pandemic on the global economy. This approach allows more effective predictions during pandemic and it will help with lowering the negative impact of COVID-19 on financial market by providing experts with additional information and tools in their decision making. Integration of SWT with BDLSTM should help not only in the current situation but also in the future situations similar to COVID-19 in order to be able to react in time and prevent a financial crisis.

First, original data of each dataset will be used as input variable in order to forecast future price movement by utilizing BDLSTM. Second, commodity and each of stock market index data will be decomposed by using SWT in order to obtain approximation and detail coefficients which will be used to train the BDLSTM model. Afterwards, the obtained results for each configuration system will be compared. Third, the impact of confirmed cases detail coefficients on forecasting accuracy will be examined. Lastly, the best performing system configuration will be used in order to show the forecasted movement of Crude Oil price for the next five days with 128 observation days. The overview of the proposed system is given in Figure 1.

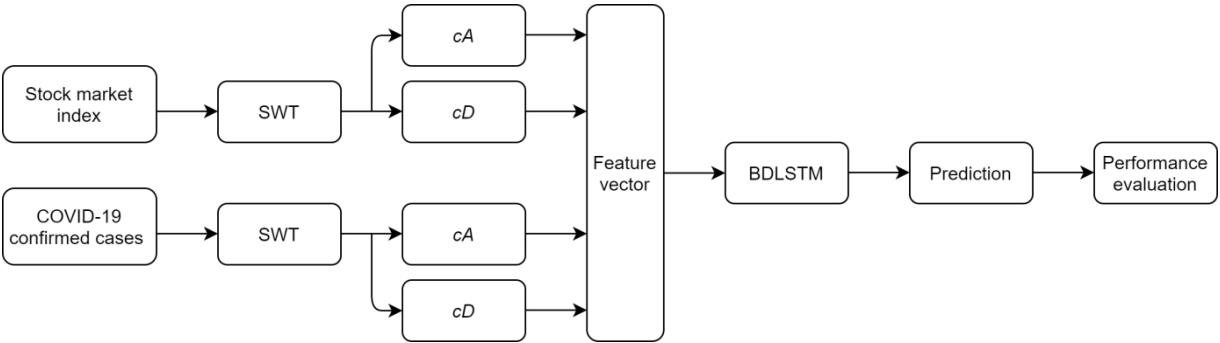

*Figure 1. Framework of the proposed system for commodity and stock price prediction during COVID-19 pandemic (SWT – Stationary Wavelet Transform, cA – Approximation coefficients, cD – Detail coefficients, BDLSTM – Bidirectional Long Short-Term Memory network)*

## 2. Materials and Methods

This section provides a detailed description of datasets used for forecasting price movements as well as a brief overview and mathematical description of Stationary Wavelet Transform, Bidirectional Recurrent Neural Network and Bidirectional Long Short-Term Memory network. In last two subsections, grid search algorithm and evaluation criteria are described.

### 2.1. Dataset description

In order to create the dataset used in this research, historical data of West Texas Intermediate (WTI) Crude Oil price and three stock indexes along with the number of COVID-19 confirmed cases are used. WTI can be defined as the main oil benchmark for North America and moreover is the most liquid crude oil benchmark [24]. In the oil market, benchmarks serve as a pricing reference for Crude Oil. Existence of different Crude Oil grades and varieties led to the use of benchmarks as a gauge in order to compare one type of Crude Oil with others. WTI is considered as light sweet oil since it has a sulfur content of 0.24%, therefore it is ideal for gasoline [25].

The stock indexes are Dow Jones Industrial Average, S&P 500, and NASDAQ Composite. The data of these indexes and Crude Oil price for the time period from March 22, 2000 to April 07, 2020 are publicly available and obtained from the Yahoo finance website [21]. Crude Oil commodity and each stock market index contain the data of volume and open, high, low, close prices for each day when the financial market was open. For the purpose of this research, only the closing price is used. The data of COVID-19 confirmed cases is publicly available and operated by the Johns Hopkins University Center for Systems Science and Engineering (JHU CSSE) and supported by ESRI Living Atlas Team and the Johns Hopkins University Applied Physics Lab (JHU APL) [22]. Obtained data contains the number of confirmed cases (infected patients) for each day since the start of the COVID-19 pandemic January 22, 2020 until April 07, 2020. Datasets are organized in a way that column represents closing price, while rows represent the date of data collection. Furthermore, for each date after January 22, 2020, the number of confirmed cases is added in additional column. Datasets with closing prices of aforementioned indexes, Crude Oil and COVID-19 confirmed cases are organized as multivariate time-series data and used in order to build an efficient deep learning model. Before implementation of the AI algorithms, signal decomposition using Wavelet Transform (WT) was utilized.

Descriptive Statistics of commodity, stock market indexes and COVID-19 confirmed cases are provided in Table 1. With these statistics, the features of each dataset can be described [26]. Descriptive Statistics used in this research are: mean, maximum, minimum, standard deviation, kurtosis and skewness. The total number of data-points i.e. observations in each of the aforementioned datasets is 4992, which were split into two parts. First part (80% of the total number) is used for model training, while the second part (20% of the total number) is used in order to evaluate the performance of the trained models.

Table 1. Descriptive Statistics of commodity, stock market indexes and COVID-19 confirmed cases.

| Statistic | Commodity | Stock market indexes | | | COVID-19 confirmed cases |
|---|---|---|---|---|---|
| | Crude Oil | DJI | S&P 500 | NASDAQ Composite | |
| Mean | 62.04356 | 14240.89 | 1596.151 | 3528.579 | 2665.000 |
| Maximum | 145.1800 | 29551.42 | 3386.150 | 9817.180 | 1430981 |
| Minimum | 17.45000 | 6547.050 | 676.5300 | 1114.110 | 0 |
| St.Dev. | 26.05498 | 5376.658 | 609.2196 | 1989.950 | 44701.96 |
| Kurtosis | 2.310495 | 3.163203 | 2.912266 | 3.200213 | 601.9265 |
| Skewness | 0.382736 | 1.104382 | 0.980250 | 1.127060 | 23.19243 |
| Observation | 4992 | 4992 | 4992 | 4992 | 4992 |

Additionally, each dataset is tested for stationarity using Augmented Dickey-Fuller (ADF) and Phillips-Perron (PP) Unit Root Tests. The results for Level and $1^{st}$ Difference are obtained with intercept and with trend and intercept for both ADF and PP tests as shown in Table 2. In order to select optimal lag length in the ADF test, the Schwarz Information Criterion (SIC) was utilized with maximum lags of 31. On the other hand, in the PP test Bartlett kernel was used as Spectral estimation method along with the Newey-West automatic bandwidth selection. The value of optimal lag length (ADF test) and optimal bandwidth (PP test) for each dataset is enclosed in parentheses () and given in Table 2. The critical values for ADF and PP tests with intercept are: -3.431479, -2.861924 and -2.567017 for 1%, 5% and 10%, while the critical values for the same tests but with trend and intercept are: -3.959877, -3.410705 and -3.127138 for 1%, 5% and 10%.

Table 2. Results of Augmented Dickey-Fuller and Phillips-Perron Unit Root Tests for each stock market index (Dow Jones Industrial Average, S&P 500 and NASDAQ Composite), Crude Oil price and COVID-19 confirmed cases.

| Variables | ADF test (with intercept) | | ADF test (with trend and intercept) | | PP test (with intercept) | | PP test (with trend and intercept) | |
|---|---|---|---|---|---|---|---|---|
| | Level | 1st Difference | Level | 1st Difference | Level | 1st Difference | Level | 1st Difference |
| Crude Oil | -1.813368 (1) | -74.18251 (0) | -1.451348 (1) | -74.19782 (0) | -1.851205 (10) | -74.15676 (10) | -1.487574 (9) | -74.17632 (10) |
| DJI | -0.589078 (9) | -21.95618 (8) | -2.502465 (9) | -21.96959 (8) | -0.561936 (22) | -82.15463 (21) | -2.445644 (23) | -82.15622 (20) |
| S&P 500 | -0.300921 (9) | -22.33898 (8) | -2.541671 (9) | -22.38022 (8) | -0.304412 (20) | -82.90155 (19) | -2.578981 (20) | -82.96314 (18) |
| NASDAQ Composite | 0.340195 (9) | -22.89305 (8) | -2.983702 (9) | -23.02818 (8) | 0.172873 (17) | -81.56604 (16) | -3.619394 (16) | -81.75234 (14) |
| COVID 19 cases | 40.89298 (30) | 51.75262 (31) | 41.05363 (30) | 51.64084 (31) | 264.3043 (42) | -34.20378 (34) | 262.4940 (42) | -34.44749 (34) |

From the results of Unit Root Tests, it can be concluded that the series of commodity and three U.S. stock market indexes do not reject the null hypothesis and can be considered as non-stationary at the level except for NASDAQ Composite, where PP test with trend and intercept shows the value of -3.619394. If test critical value of 5% (-3.410705) is chosen, the null hypothesis can be rejected, and the series of NASDAQ Composite is stationary. Furthermore, results point out that commodity and three stock market indexes are stationary at their 1st difference form.

In the case of COVID-19 confirmed cases, the results reveal that the series reject the null hypothesis in the PP test with intercept and with trend and intercept at the 1st difference and can be considered as stationary.

### 2.2. Data decomposition with Wavelet Transform

The Wavelet Transform (WT) is a powerful mathematical tool for signal processing [27]. Applying WT, a signal can be decomposed into many frequency bands, which can simplify the analysis process. The Fourier Transform (FT) major drawback is losing time information, while preciseness of Short Time Fourier Transform (STFT) largely depends on its window size and shape. Unlike FT and STFT, WT preserves precise information about time and frequency. Since the characteristics of the stock market are non-stationary, non-linear and noisy, and considering the aforementioned drawbacks of FT and STFT, WT can be appropriate approach when dealing with economic and financial time-series analysis. Wavelet transform of signal $x(t)$ can be calculated as:

$$X(\tau, a) = \frac{1}{\sqrt{|a|}} \int_{-\infty}^{\infty} x(t)\psi^* \left(\frac{t-\tau}{a}\right) dt, \qquad (1)$$

where $\psi$ represents the analyzing wavelet, * stands for complex conjugate, $a$ represents a time dilation and $\tau$ represents time translation [28]. Therefore, the Discrete Wavelet Transform (DWT) of signal $x[m]$ can be defined as [29]:

$$X[k, l] = 2^{-\frac{k}{l}} \sum_{m=-\infty}^{\infty} x[m]\psi[2^{-k}m - l]. \qquad (2)$$

To obtain approximation coefficients *cA* and detail coefficients *cD* from the original signal *x*[*m*], DWT needs to be performed. After DWT decomposition process, approximation contains the low-frequency components while the detail contains the high-frequency components of the original signal. In the case of conventional DWT, after each decomposition level signal is decimated. Because of decimation, the DWT is not a time-invariant transform and it is not suitable for data preprocessing in this research. This drawback can be overcome by using one of the DWT's extensions, such as, Stationary Wavelet Transform (SWT) which solves the problem of shift-invariance. SWT is feasible for feature extraction, change detection and pattern recognition due to shift-invariant and non-decimated properties [30]. In SWT, after the signal is convolved with high and low pass filters, no decimation is performed, thereby the number of obtained coefficients *cA* and *cD* at each decomposition level is the same as the number of samples in the original signal. Five-level SWT decomposition of an input signal *x*[*n*] is shown in Figure 2.

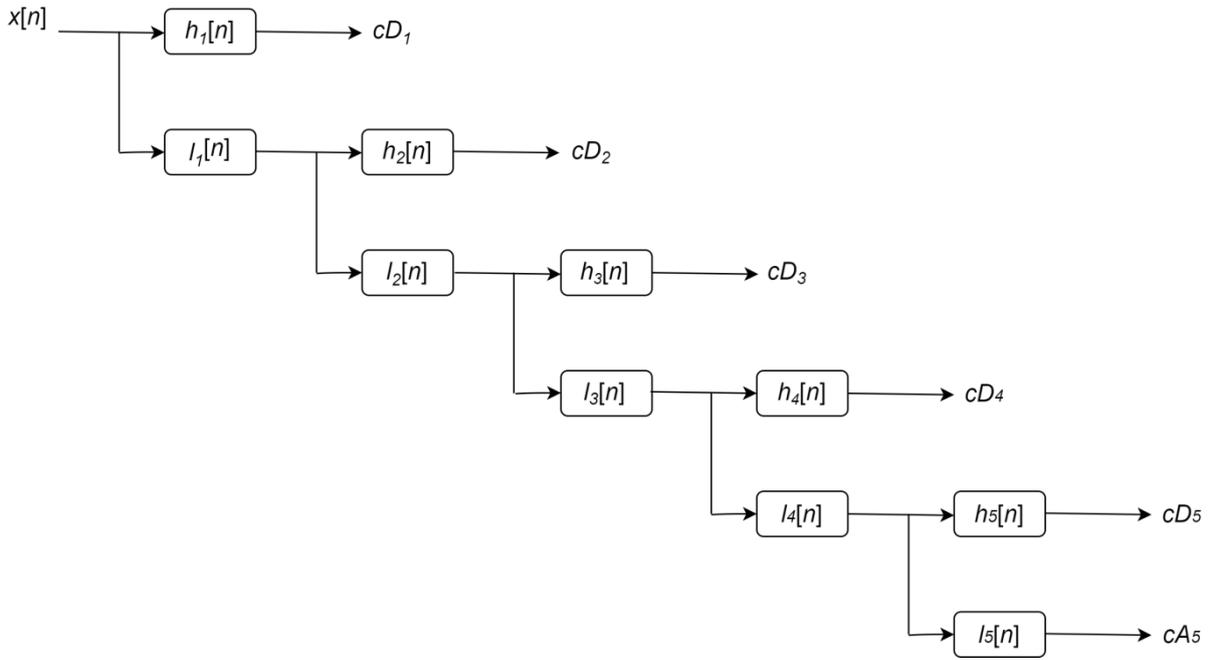

*Figure 2. Signal decomposition using SWT at level five ($h_i$ – high level features at decomposition level i, $l_i$ – low level features at decomposition level i, $cD_i$ – Detail coefficients at decomposition level i, $cA_5$ – Approximation coefficients at decomposition level five)*

In order to obtain a good decomposition of the original signal, discrete Meyer wavelet is utilized. The Meyer wavelet is linear-phase, orthogonal wavelet, and it is defined in the frequency domain as follows [31]:

$$\psi(\omega) = \begin{cases} \frac{1}{\sqrt{2\pi}} \sin\left(\frac{\pi}{2} v\left(\frac{3|\omega|}{2\pi} - 1\right)\right) e^{j\omega/2} & if\ \frac{2\pi}{3} < |\omega| < \frac{4\pi}{3}, \\ \frac{1}{\sqrt{2\pi}} \cos\left(\frac{\pi}{2} v\left(\frac{3|\omega|}{4\pi} - 1\right)\right) e^{j\omega/2} & if\ \frac{4\pi}{3} < |\omega| < \frac{8\pi}{3}, \\ 0 & otherwise, \end{cases} \qquad (3)$$

where *v* is an auxiliary function that can be defined as

$$v(a) = a^4(35 - 84a + 70a^2 - 20a^3), \quad a \in [0, 1]. \qquad (4)$$

### 2.3. Bidirectional Recurrent Neural Network

Recurrent Neural Networks (RNNs) are a class of Artificial Neural Networks (ANNs) with feedback connection [32]. Between units are connections by which a directed cycle is formed. Therefore in RNN model, a signal can travel both forward and backward. In such network, knowledge can be represented with the values of synaptic connections between input, hidden and output layers of neurons. The main idea behind RNNs is to use sequential data as input. The RNN model can be simplified by unfolding the RNN architecture over the input sequence of data as is shown in Figure 3.

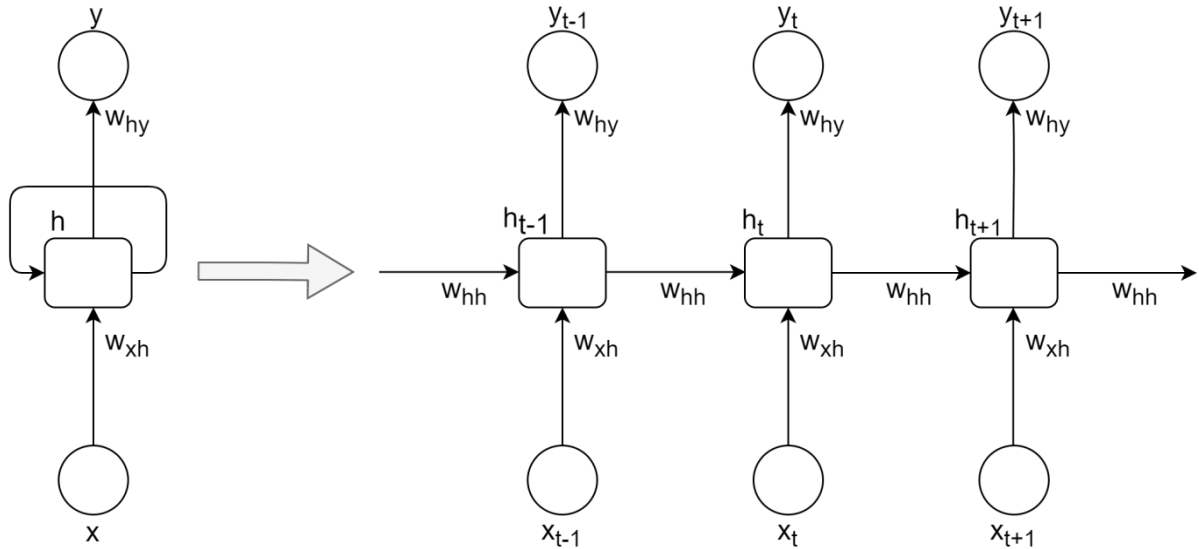

*Figure 3. Folded RNN architecture and the process of unfolding with T time-steps (x – input vector, w$_{xh}$ – weight matrix between input and hidden layer, h – hidden state, w$_{hh}$ – weight matrix between two hidden states, w$_{hy}$ – weight matrix between hidden and output layer, y – output vector)*

Conventional feedback neural networks process the data in one direction only, but in certain areas, past and future information is desirable. Therefore, in 1997, Schuster and Paliwal introduced Bidirectional Recurrent Neural Network (BRNN), whose basic idea was to extend the RNN architecture by introducing additional hidden layers where data were placed in the opposite, negative direction. The hidden layer maintains a hidden state which can be defined as:

$$\vec{h}_t = \sigma(W_{x\vec{h}}x_t + W_{\vec{h}\vec{h}}\vec{h}_{t-1} + b_{\vec{h}}) \qquad (5)$$

for the positive direction, and

$$\overleftarrow{h}_t = \sigma(W_{x\overleftarrow{h}}x_t + W_{\overleftarrow{h}\overleftarrow{h}}\overleftarrow{h}_{t+1} + b_{\overleftarrow{h}}) \qquad (6)$$

for the negative direction [33]. $W_{xh}$ represents the weight matrix between input and hidden layer, $x_t$ represents the input vector, $W_{hh}$ represents the weight matrix between two hidden states, $b_h$ represents the bias of the hidden layer and $\sigma$ represents the activation function. The output layer can be defined as:

$$y_t = \sigma\left(W_{\vec{h}y}\vec{h}_t + W_{\overleftarrow{h}y}\overleftarrow{h}_t + b_y\right) \quad (7)$$

where $W_{\vec{h}y}$ represents the weight matrix between hidden and output layer, while $W_{\overleftarrow{h}y}$ represents the same but in other direction, $b_y$ is the bias of the output layer [33]. As a major drawback, BRNN in its basic form cannot model a complex time dynamics and it can suffer from the vanishing or exploding gradients.

### 2.3. Bidirectional Long Short-Term Memory

One of the solutions to overcome aforementioned problems is to use Bidirectional Long Short-Term Memory (BDLSTM) architecture. Such architecture differs from the RNN architecture in terms of hidden layers. BDLSTM has a LSTM cell as hidden layer, which consists of three gates: an input gate, forget gate, and an output gate. LSTM cell can be mathematically defined as follows [34]:

$$f_t = \sigma(W_f x_t + U_f h_{t-1} + b_f), \quad (8)$$

$$i_t = \sigma(W_i x_t + U_i h_{t-1} + b_i), \quad (9)$$

$$o_t = \sigma(W_o x_t + U_o h_{t-1} + b_o), \quad (10)$$

$$C_t = f_t \odot C_{t-1} + i_t \odot \tanh(W_c x_t + U_c h_{t-1} + b_c) \text{ and} \quad (11)$$

$$h_t = o_t \odot \tan h(C_t). \quad (12)$$

In Eq. (8) – (12) $f_t, i_t, o_t$ represent forget, input, and output gate, $W$ and $U$ represent the weight matrices, $b$ is a bias vector, $\sigma_g$ is a sigmoid activation function, *tanh* is the hyperbolic tangent function, $C_t$ is the cell output state, $h_t$ is the layer output, and operator $\odot$ is the element-wise product of the vectors. By using Eq. (5) – (11), forward and backward layer outputs can be calculated. The result of combining BRNN with LSTM cells is a BDLSTM network, which can model more complex time dynamics and deal with long-term dependencies [35]. The architecture of an unfolded BDLSTM is shown in Figure 4.

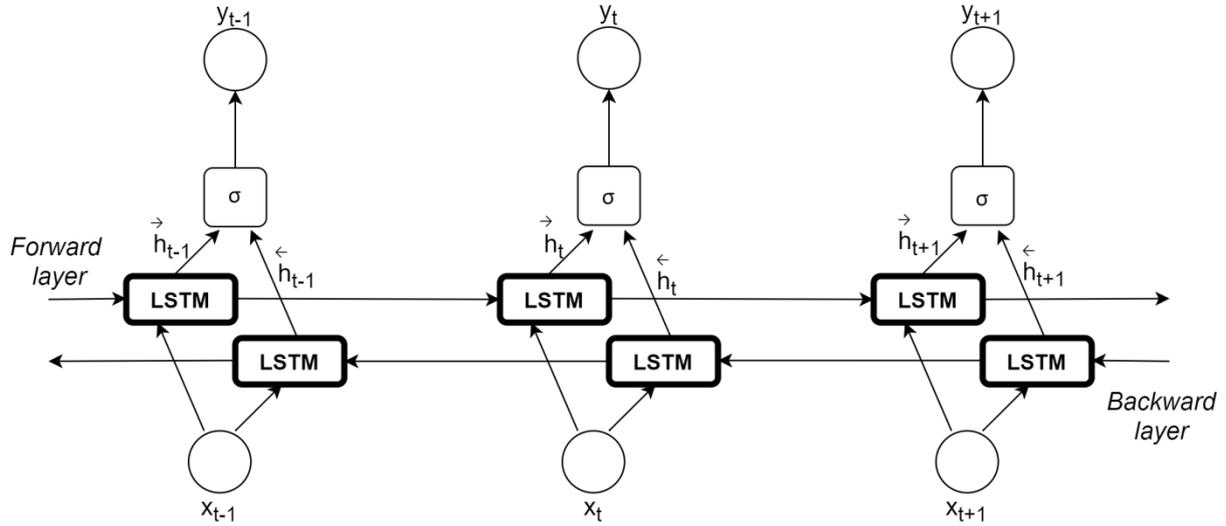

*Figure 4. The architecture of an unfolded BDLSTM with three steps (x – input vector, $\vec{h}$ – hidden state for the positive direction, $\overleftarrow{h}$ – hidden state for the negative direction, σ – function used to combine two output sequences, y – output vector)*

By using inputs in a positive sequence, the forward layer output sequence is calculated, and by using reversed inputs, the backward layer output sequence is calculated. Each element in output vector of BDLSTM layer can be calculated as:

$$y_t = \sigma(\vec{h_t}, \overleftarrow{h_t}), \quad (13)$$

where two output sequences are combined by utilizing the σ function [35]. In many studies, bidirectional networks have been proven to be significantly better than unidirectional networks in various fields, such as speech recognition [36], classification problems [37], and also in stock price prediction [38]. In this research, BDLSTM is trained in order to predict price movement for the time period where the impact of COVID-19 on the global economy is relatively high. In the output vector of a BDLSTM layer, the last element is predicted value for the next time iteration. Furthermore, to prevent the network from overfitting, dropout can be implemented on hidden layers [39].

## 2.4. Hyperparameter Optimization

In order to determine optimal hyperparameters of the ANN, the grid search algorithm has been used. This algorithm can be described as an exhaustive search through a set of manually specified parameters [40]. Therefore it iterates through every possible parameter combination, trains the network and finally stores the result for each combination. Hyperparameters can be described as follows [41]:

- hidden layer size is defined with two integers where first one represents the number of hidden layers and the other one defines the number of hidden neurons in that layer,
- activation function determines the output value behavior of each neuron based on its input values,
- optimizer is used for minimizing the value of cost function in order to improve metric important for the research,
- learning rate can be considered as a hyperparameter that regulates the weight adjustment,
- learning rate decay is a technique where the training process starts with a large learning rate, and then decays it with the time, and
- regularization parameter L2 forces the weights to decay towards zero but does not make them zero in order to limit the influence of input parameters.

This way the algorithm can find the optimal hyperparameters of the model that achieve the most accurate predictions. The hyperparameters adjusted in this research are: number of BDLSTM hidden layers and neurons, number of fully-connected (FC) hidden layers, activation function, optimizer, learning rate, learning rate decay, and regularization parameter L2. Subset of hyperparameter space is shown in Table 3.

Table 3. Hyperparameter values used in model training process. First column represents hyperparameter name, and in the second column possible parameters are shown.

| Hyperparameter | Possible parameters |
|---|---|
| BDLSTM - hidden layer size | (32), (64), (32, 32), (64, 32), (64, 64), (32, 32, 32), (64, 32, 32), (64, 32, 64), (64, 64, 64) |
| FC – hidden layer size | (12), (24), (12, 12), (24, 12), (24, 24), (12, 12, 12), (24, 12, 12), (24, 12, 24), (24, 24, 24) |
| Activation function | ReLU, ELU, Tanh, Identity |
| Optimizer | Adam, RMSprop |
| Learning rate | 0.0001, 0.001, 0.01 |
| Learning rate decay | 1e-7, 1e-6, 1e-5 |
| Regularization parameter - L2 | 0.0001, 0.001, 0.01 |

**2.5. Evaluation criteria**

In order to evaluate the performance of the implemented model, two evaluation criteria can be used as accuracy measures. These performance measures are Mean Absolute Error (MAE) and Root Mean Square Error (RMSE), and can be calculated as follows [42]:

$$MAE = \frac{1}{N}\sum_{t=1}^{N}|y_t - \hat{y}_t| \text{ and} \qquad (14)$$

$$RMSE = \sqrt{\frac{1}{N}\sum_{t=1}^{N}(y_t - \hat{y}_t)^2}, \qquad (15)$$

with $y_t$ being the true signal, and $\hat{y}_t$ being forecasted signal. Smaller values of performance measures defined by Eq. (14) and Eq. (15) mean the better forecasting performance of the model and vice-versa.

## 3. Results

The forecasting results are obtained for Crude Oil commodity and Dow Jones Industrial Average, S&P 500, NASDAQ Composite indexes. For each dataset, SWT is performed in order to obtain their approximation and detail coefficients at five decomposition levels using discrete Meyer wavelet function. For example, such decomposed signal of Crude Oil price is shown in Figure 5. where *s* is stock closing price for time period from March 22, 2000 to April 07, 2020, *cA* and *cD* are approximation and detail coefficients.

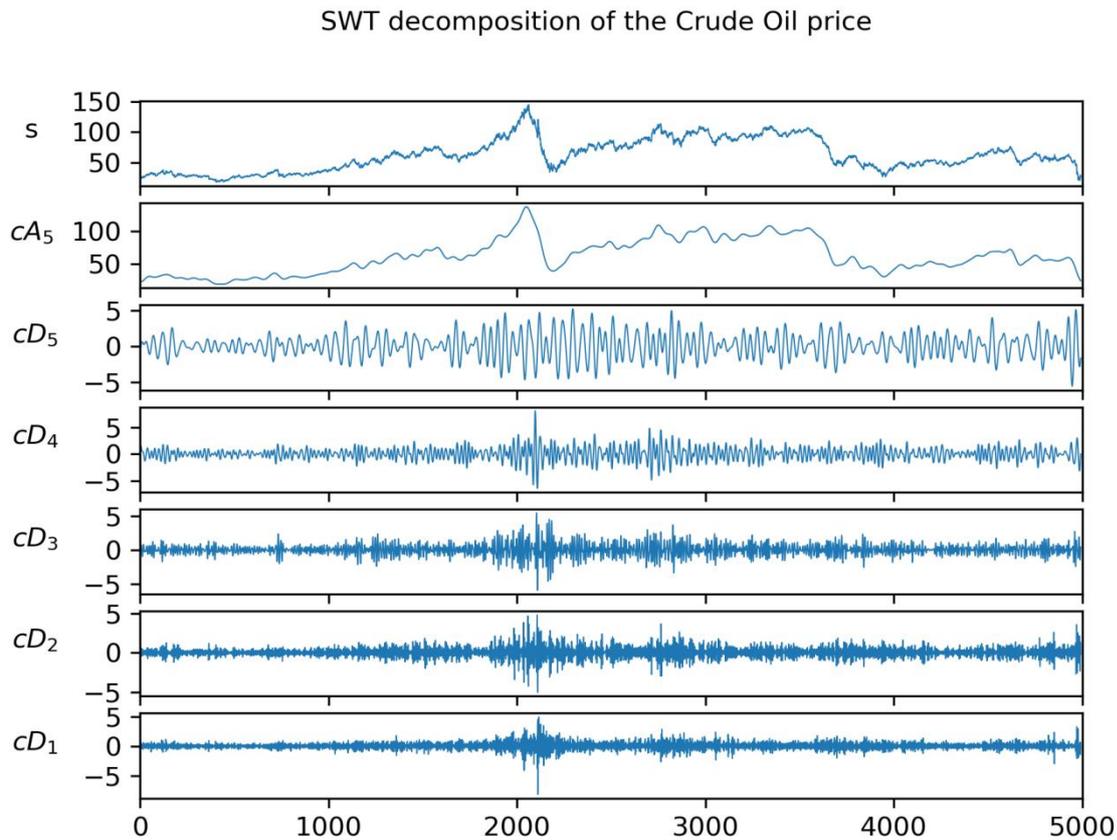

*Figure 5. Five-level decomposition of Crude Oil closing price using SWT (s – input data, $cA_i$ – Approximation coefficients at decomposition level i, $cD_i$ – Detail coefficients at decomposition level i)*

Three main system configurations were examined in order to achieve high-quality regression and small values of performance measures. In the first configuration, non-preprocessed data is used to train the BDLSTM model, in second the BDLSTM model is trained by using both approximation and detail coefficients (AD). Finally in the last configuration, the data contain approximation and detail coefficients for commodity and stock index price, but only the approximation for COVID-19 confirmed cases (ADA). The values of performance measures for Crude Oil and stock market indexes with system configurations are shown in Table 4.

Table 4. Simulation results and performance comparison of three configuration systems for Crude Oil commodity and Dow Jones Industrial Average, S&P 500, NASDAQ Composite indexes. As evaluation criteria, RMSE and MAE are used.

|  |  | RMSE | MAE |
|---|---|---|---|
| **CRUDE OIL** | BDLSTM | 0.04083 | 0.03051 |
|  | BDLSTM+WT-AD | 0.03685 | 0.02962 |
|  | BDLSTM+WT-ADA | 0.02911 | 0.02315 |
| **DJI** | BDLSTM | 0.04557 | 0.02753 |
|  | BDLSTM+WT-AD | 0.02436 | 0.01743 |
|  | BDLSTM+WT-ADA | 0.01450 | 0.01014 |
| **S&P 500** | BDLSTM | 0.04391 | 0.02627 |
|  | BDLSTM+WT-AD | 0.03670 | 0.02883 |
|  | BDLSTM+WT-ADA | 0.02389 | 0.01734 |
| **NASDAQ** | BDLSTM | 0.03632 | 0.02502 |
|  | BDLSTM+WT-AD | 0.02610 | 0.01754 |
|  | BDLSTM+WT-ADA | 0.02597 | 0.01339 |

BDLSTM model that achieves the best results have the same architecture for commodity and stock market indexes. Such architecture consists of three hidden layers, where the first two are BDLSTM layers with 64 hidden neurons each and the last one is FC layer with 12 hidden neurons. Additionally, dropout is applied on BDLSTM hidden layers with the value of 0.2 for the first, and 0.1 for the second layer. All of the hidden layers use *tanh* activation function, and *Adam* optimizer. The best model has a learning rate of 0.001, learning rate decay of 1e-6 and a regularization parameter of 0.0001.

During the COVID-19 pandemic, the correlation between Crude Oil price and other stock market indexes used in this research exist, and all of the data can be rearranged and used as multivariate time-series data. This way, important features of Crude Oil commodity and three stock market indexes can be captured in order to predict movement of one price more precisely.

By using data of Crude Oil commodity, three stock indexes and information of COVID-19 confirmed cases in the past 128 days, predictions were made for Crude Oil price for the next five days, as shown in Figure 6. The values of performance measures for Crude Oil with BDLSTM+WT-ADA system configuration are shown in Table 5.

Table 5. Simulation results obtained with the best configuration system for Crude Oil price. As evaluation criteria, RMSE and MAE are used.

|  |  | RMSE | MAE |
|---|---|---|---|
| **CRUDE OIL** | BDLSTM+WT-ADA | 0.02116 | 0.01701 |

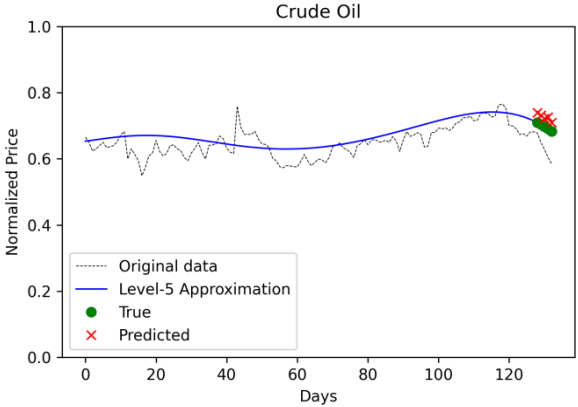

(a) Time period July 16, 2019 – January 27, 2020

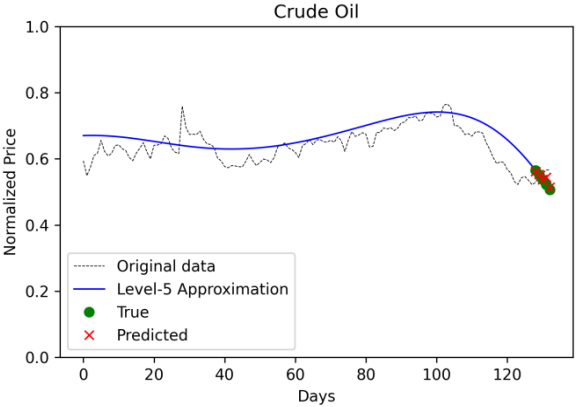

(b) Time period August 06, 2019 – February 18, 2020

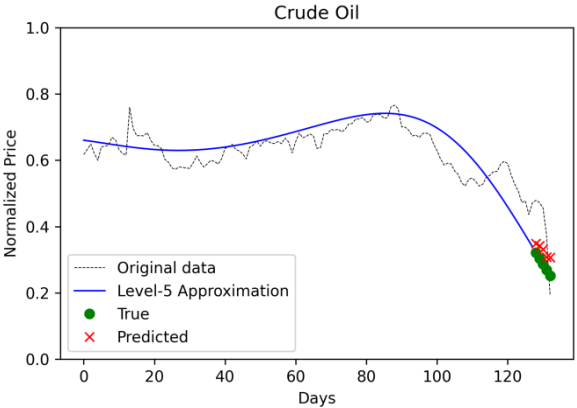

(c) Time period August 27, 2019 – March 10, 2020

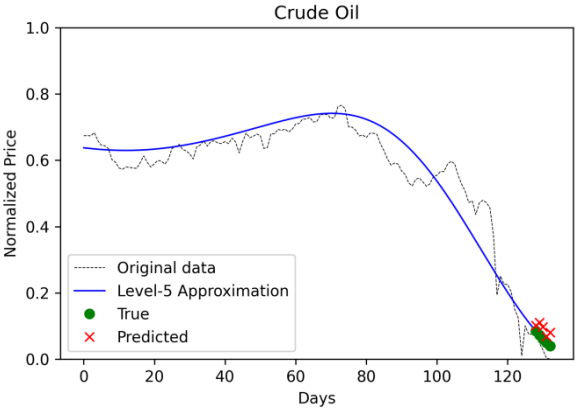

(d) Time period September 18, 2019 – March 31, 2020

*Figure 6. Prediction of Crude Oil price movement based on observations in the past 128 days*

## 4. Discussion

This research proposes an integrated system, BDLSTM+WT-ADA, for commodity and stock price movement prediction during current pandemic. In order to validate the feasibility, the proposed system is compared with other approaches presented in the literature [43, 44] for prediction of stock prices. When all results are summed up, it can be seen that the minimal values of RMSE and MAE are achieved using BDLSTM+WD-ADA system configuration. If forecasting performances of three system configurations are compared, it can be seen that all of the configurations achieved RMSE value of 0.04557 or smaller and MAE value 0.03051 or smaller. These results are satisfactory in terms of forecasting commodity and stock market price. Furthermore, it can be seen that impact of approximation and detail coefficients manifest through simulation results. For example, the worst results for each of the stock market index and Crude Oil commodity are achieved by using original, non-preprocessed data. Moreover, the best results are achieved using the proposed system configuration BDLSTM+WD-ADA with the lowest RMSE value (0.01450) and MAE value (0.01014) for Dow Jones Industrial Average index. Such configuration (ADA) uses five-level decomposition utilizing the SWT with discrete Meyer wavelet function.

Crude Oil is globally the most important commodity and is driven by supply and demand as any other good, but has a tendency to fluctuate more in price than, for example, stocks and bonds on financial markets. As Crude Oil prices rise, so do other fuel prices, which increase production prices in general. Rising production prices lead to higher prices of food and industrial products thus generating inflation. The reduced demand for Crude Oil caused by various impacts, in this case the global pandemic, results in Crude Oil price disruption and, as mentioned, has a profound effect on the economy in general. For this reason, Crude Oil price was selected for five-day prediction that can be extremely useful for foreseeing the events that follow.

The relationship between the COVID-19 confirmed cases and the crude oil price is significant. With an increase in the number of cases, measures are being taken to slow down further spread. Some of them are closing factories, offices and shops and restricting the movement. Consequently, much less fuel is needed for vehicles, machinery, etc. If demand decreases and supply remains unchanged, this leads to lower commodity prices and crude oil prices fall [45]. The same goes for the stock market. If companies on the stock market reduce or close their operations, shareholders become nervous and fear what will happen to the value of that company's shares in the future and whether it will decline. They start selling stocks thus increasing the supply in the market. As the number of confirmed cases increases and measures become more stringent, other buyers are not interested in buying. If there are more participants in the market looking to sell a stock than there is demand to acquire the stock, the stock price will fall. Therefore, the inclusion of a large amount of data (confirmed cases for each day) allows us to have more accurate information and a more credible result.

From obtained results of forecasting Crude Oil price movement it can be seen that proposed system configuration is capable for accurate five-day prediction based on observations in the past 128 days. In the middle of February, the first signs of decline in Crude Oil price were observed and for that time period, BDLSTM+WD-ADA system configuration successfully predicted the price movement. Expectations about future events are extremely important in times of crisis in order to adequately respond and initiate measures and mechanisms for the preservation and stability of the economy. However, because of the global role of Crude Oil as still irreplaceable source of energy, it has the direct effect on the geopolitical trends. The price of Crude Oil is, from the economic aspect, hard to predict precisely due to political relationships in the triangle: USA-OPEC countries-Russia, which are rarely stable and will always disturb the economic model of supply and demand in a competitive market. Consequently, prediction models are valuable and can be used to foresee the sequence of events, but factors like political interference, that cannot be included in the model, also affect the price and must be emphasized.

### 4.1. Result comparison

The obtained results demonstrate the connection between the crude oil price and the number of active cases of COVID-19. Most of the research performed in the area of economic impact of COVID-19 concludes that the rising number of active COVID-19 cases has a large negative impact on global markets, as shown. Baker et al. (2020) use disaster modeling techniques which predict a GDP contraction in the USA – with as much as 20 percent contraction being predicted with 90 percent confidence interval [46]. Toda (2020) shows the possibility of a temporary 50 percent stock price decrease using classic asset pricing modeling [47]. Baldwin and Tomiura (2020) conclude that there is danger of permanent damage to the trade system, depending on the policies implemented [48]. Atkeson (2020) uses as SJR Markov chain based model to determine the spread and comments the possibility of key financial and economic infrastructure being affected temporarily and permanently due to possible extreme staff shortages, in case where the number of active cases exceeds 10 percent of population [49]. Albulescu (2020) investigates the impact of COVID-19 on oil pricing, due to the initial 20 percent drop caused by the market being flooded with oil [50]. AutoRegressive Distributed Lag (ARDL) estimation performed by the author demonstrates that daily new infections have only a marginal impact, but a larger indirect impact is caused due to the amplification of financial market volatility, falling in line with prediction made in this paper. McKibbin and Fernando (2020) observe seven different scenarios in regard to the global macroeconomic impact of COVID-19, concluding that even a small, contained, impact can have a large negative influence on the global markets [51]. Fernandes (2020) analyzes the reports from 30 countries under varying scenarios and concludes that the possible impact of COVID-19 on the world economy is being underestimated, especially in heavily service oriented countries [23]. Fernandes discusses that one of the possible problems is underestimation of impact due to modeling based on previous SARS infections – showing the need for newer, fast modeling techniques, which can, as shown in this and other papers be AI based [52, 53].

## 5. Conclusion

The goal of this research was to generate a forecasting model that integrates Stationary Wavelet Transform and Bidirectional Long Short-Term Memory networks in order to predict commodity and stock price movement during the COVID-19 pandemic. The results obtained using proposed BDLSTM+WT-ADA configuration system show that, in addition to tradition statistical models, Artificial Intelligence algorithms can be used to predict the movements of financial markets. The peculiarity of this paper is that information of COVID-19 confirmed cases is used as input data in parallel with three leading U.S. stock market indexes along with Crude Oil commodity. For the normal functioning of the global economy, it is very important that Crude Oil has stable and secured delivery to the market. The global economy has been slowly recovering since the financial crisis of 2007-2008, but COVID-19 outbreak already showed a huge impact on energy prices as well as stock market. Our proposed system shows a decline in Crude Oil price. In addition to predicting future events through the methods that are presented, it is important to note that the geopolitical aspect is indirectly included in presented model through the input data. There-fore it is not possible to clearly define the impact of geopolitical aspects in here presented model. It can be assumed that the geopolitical aspect in this model is negligible, but it has a significant impact on the global economy.

The observed period used in analysis was marked by the extreme increase in oil stocks on the market. Due to this over-supply from the most important exporting countries (e.g. OPEC countries) and geopolitical issues between the major players on the market, prices were consequently slumping. Following the trends after the research was conducted, it is concluded that despite the increase in the number of COVID-19 confirmed cases, the market is gradually adjusting oil prices due to the fact of joint agreement on production cuts (lowering the supply side) and on the other hand the gradual opening of markets and recovery of demand. The logical consequence is the growing demand on a global level simultaneously followed by the improvement of relations between oil exporters, which contributes to the temporary market stability.

Unexpected situations such as a pandemic can have a significant effect on market fundamentals in the short term, and there have been correlations with indexes and oil. Due to further observation, in a period of several months and through the gradual opening of economies, there is a stabilization of supply and demand, which has a positive effect on the formation of market equilibrium. The continued movement of stock indexes, especially this positive movement, does not reflect the real situation in the economy but is primarily based on expectations and is further stimulated by monetary and fiscal incentives (e.g. cut of interest rates, reduction of taxes) from national governments.

The main contribution and novelty of the presented research is not only demonstrating the existence of a link between the COVID-19 infections and commodity prices along with stock market prices but showing that modeling of the same can be achieved using data driven, artificial based, modeling methods.

Future work should use datasets with more data-points i.e. long time historical intraday data in order to achieve more precise forecasting. Also, apply more AI algorithms such as Dynamic Programming (DP), Genetic Programming (GP) and combination of Convolutional Neural Networks (CNNs) with LSTM network in attempt to find more robust systems. The main idea of using such algorithms will be to develop an advanced automatic forecasting system with capability of recognizing the positive correlation between financial markets.

## Data Availability

This research uses a publicly available financial market data published by Yahoo finance website and a publicly available dataset "2019 Novel Coronavirus Data Repository" published by Johns Hopkins University Center for Systems Science and Engineering (JHU CSSE).

## Conflicts of Interest

The authors declare that there is no conflict of interest regarding the publication of this paper.

## Funding Statement

This research has been (partly) supported by the CEEPUS network CIII-HR-0108, European Regional Development Fund under the grant KK.01.1.1.01.0009 (DATACROSS), project CEKOM under the grant KK.01.2.2.03.0004 and University of Rijeka scientific grant uniri-tehnic-18-275-1447.